\documentclass[usenatbib]{mn2e}
\usepackage{graphicx,natbib,amssymb,amsmath}
\usepackage{txfonts}
\usepackage{color}

\newcommand{\Teb}{{\bar{T}_{\rm e}}}
\newcommand{\Teby}{{\bar{T}^y_{\rm e}}}
\newcommand{\Tebell}{{\bar{T}^{yy}_{\rm e, \ell}}}

\newcommand{\keV}{{\rm keV}}
\newcommand{\GHz}{{\rm GHz}}
\newcommand{\Te}{T_{\rm e}}
\newcommand{\pot}[2]{#1 \times 10^{#2}}
\newcommand{\beal}{\begin{align}}
\newcommand{\bsub}{\begin{subequations}}
\newcommand{\esub}{\end{subequations}}

\begin{document}

\title[Relativistic corrections in tSZ analyses]
{Can we neglect relativistic temperature corrections in the Planck thermal SZ analysis?}

\author[Remazeilles et al.]{Mathieu Remazeilles$^1$\thanks{E-mail:mathieu.remazeilles@manchester.ac.uk}, Boris Bolliet$^1$\thanks{E-mail:boris.bolliet@manchester.ac.uk}, Aditya Rotti$^1$\thanks{E-mail:aditya.rotti@manchester.ac.uk} and Jens Chluba$^1$\thanks{E-mail:jens.chluba@manchester.ac.uk}
\\
$^1$Jodrell Bank Centre for Astrophysics, School of Physics and Astronomy, The University of Manchester, Manchester M13 9PL, U.K.
}

\date{\vspace{-0mm}{Accepted 2018 --. Received 2018 --}}

\maketitle

\begin{abstract}
Measurements of the thermal Sunyaev-Zel'dovich (tSZ) effect have long been recognized as a powerful cosmological probe. 
Here we assess the importance of relativistic temperature corrections to the tSZ signal on the power spectrum analysis of the {\it Planck} Compton-$y$ map, developing a novel formalism to account for the associated effects. The amplitude of the tSZ power spectrum is found to be sensitive to the effective electron temperature, $\Teb$, of the cluster sample. 
Omitting the corresponding modifications leads to an underestimation of the $yy$-power spectrum amplitude. Relativistic corrections thus add to the error budget of tSZ power spectrum observables such as $\sigma_8$. This could help alleviate the tension between various cosmological probes, with the correction scaling as $\Delta \sigma_8/\sigma_8 \simeq 0.019\,[k\Teb/5\,\keV]$ for {\it Planck}. At the current level of precision, this implies a systematic shift by $\simeq 1\sigma$, which can also be interpreted as an overestimation of the hydrostatic mass bias by $\Delta b \simeq 0.046\,(1-b)\,[k\Teb/5\,\keV]$, bringing it into better agreement with hydrodynamical simulations. It is thus time to consider relativistic temperature corrections in the processing of current and future tSZ data.
\end{abstract}


\section{Introduction}
The thermal Sunyaev-Zel'dovich (tSZ) effect is now routinely used to detect clusters of galaxies \citep{STA11, Planck2013SZ}. More than $10^3$ clusters have been seen through this effect and the number of Sunyaev-Zel'dovich (SZ) clusters is expected to increase by more than one order of magnitude with future experiments \citep[e.g.,][]{Melin2018JCAP, SOWP2018}. The tSZ effect is caused by the upscattering of cosmic microwave background (CMB) photons by thermal electrons residing in the potential wells of clusters, yielding a Compton-$y$ distortion, which in the non-relativistic limit has the frequency dependence (in intensity) ${Y_0(\nu) =  (2h/c^2) (kT_{\rm CMB}/h)^3\,x^4 {\rm e}^x/({\rm e}^x-1)^2 [x\,{\rm coth}(x/2)\,-\,4}]$ \citep{Zeldovich1969, Sunyaev1980}. Here, $c$ denotes the speed of light and ${x\equiv h\nu / kT_{\rm CMB}}$ with $h$ being the Planck constant, $k$ the Boltzmann constant, and $T_{\rm CMB}$ the CMB blackbody temperature. 

The importance of SZ clusters as a cosmological probe has long been recognized \citep[e.g.,][]{Sunyaev1980, Rephaeli1995ARAA, Birkinshaw1999, Carlstrom2002}. As the largest gravitationally bound systems, clusters are a unique tracer of the large-scale structure in the Universe. Multifrequency observations with the {\it Planck} satellite allow us to extract valuable information about the distribution of matter on the largest scales. One example is the large-scale lensing potential, which  was mapped for the first time with {\it Planck} \citep{Planck2016Lens}. Similarly, {\it Planck} revealed the first Compton-$y$ map, which through the tSZ effect informs us about the integrated electron pressure along different lines of sight \citep{Planck2016ymap}.

The clusters observed with {\it Planck} are massive and contain a hot electron plasma that is also seen in X-rays \citep{Vikhlinin2006, Leccardi2008, 2010A&A...517A..92A}. The thermal velocities of electrons inside massive clusters can be appreciable, reaching a fair fraction of the speed of light ($\varv_{\rm th}\simeq 0.1 -0.2c$). In this situation, the non-relativistic approximation for the SZ signal \citep{Zeldovich1969}, commonly used in CMB analysis, no longer suffices, and relativistic temperature corrections become important \citep{Wright1979, Fabbri1981, Rephaeli1995, Sazonov1998, Challinor1998, Itoh98}. These corrections are currently hard to detect and have been searched for in individual clusters \citep[e.g.,][]{Hansen2002, Zemcov2012, Prokhorov2012, Chluba2012moments} and through stacking analyses \citep[e.g.,][]{Hurier2016, Erler2017, Hincks2018}. Here we consider the effect of relativistic corrections on the {\it Planck} tSZ power spectrum analysis, demonstrating that they already add to the current error budget, leading to a bias in the inferred matter power spectrum amplitude, i.e., $\sigma_8$.

The power spectrum of the Compton-$y$ parameter, $C^{yy}_\ell$, connects the extracted information to the underlying cosmology \citep[e.g.,][]{Refregier2000, Komatsu:2002wc}. Its amplitude depends steeply on that of the matter power spectrum, parametrized by $\sigma_8$ \citep{Komatsu1999}. Using the halo model, one finds $C^{yy}_\ell\propto \sigma_8^{8.1}$ for the contributions of SZ clusters \citep{Planck2016ymap, Bolliet:2017lha}. Similarly, the skewness of the one-point probability distribution function (PDF) of the $y$-parameter was shown to scale as $\big<y^3\big>\propto \sigma_8^{12}$ \citep{Jose2003, Wilson2012, Bhattacharya2012}. Therefore, tSZ measurements can be used to derive constraints on $\sigma_8$ \citep{Komatsu:2002wc, Planck2016ymap, Bolliet:2017lha}, albeit with obstacles from cluster astrophysics \citep{Battaglia2010, Shaw2010SZ, Battaglia2012}, foregrounds \citep{Planck2016ymap} and systematics \citep[e.g.,][]{Planck2013components, Planck2016ymap}.

Evidently, we do not directly measure $C^{yy}_\ell$. We use multifrequency observations to obtain maps of the $y$-parameter, which then allow us to estimate $C^{yy}_\ell$. In the intermediate steps, one of the crucial approximations is that the spectral shape of the tSZ signal, $Y(\nu)$, is the same for all clusters. Thus, the tSZ power spectrum at one frequency is given by $C^{\rm tSZ}_\ell(\nu)\propto \big<Y^2(\nu)\,|y_\ell|^2\big>=Y^2(\nu) \, C^{yy}_\ell$, with $C^{yy}_\ell=\big<|y_\ell|^2\big>$.\footnote{We are hiding details of the ensemble average of the single-cluster contribution over the mass function and cosmological volume by $\big<\ldots\big>$ \citep[see][for details]{Komatsu:2002wc}.} An additional important simplification is that $Y(\nu)$ is approximated using the non-relativistic limit, $Y(\nu)\simeq Y_0(\nu)$. 

\begin{figure}
\begin{centering}
\includegraphics[width=0.98\columnwidth]{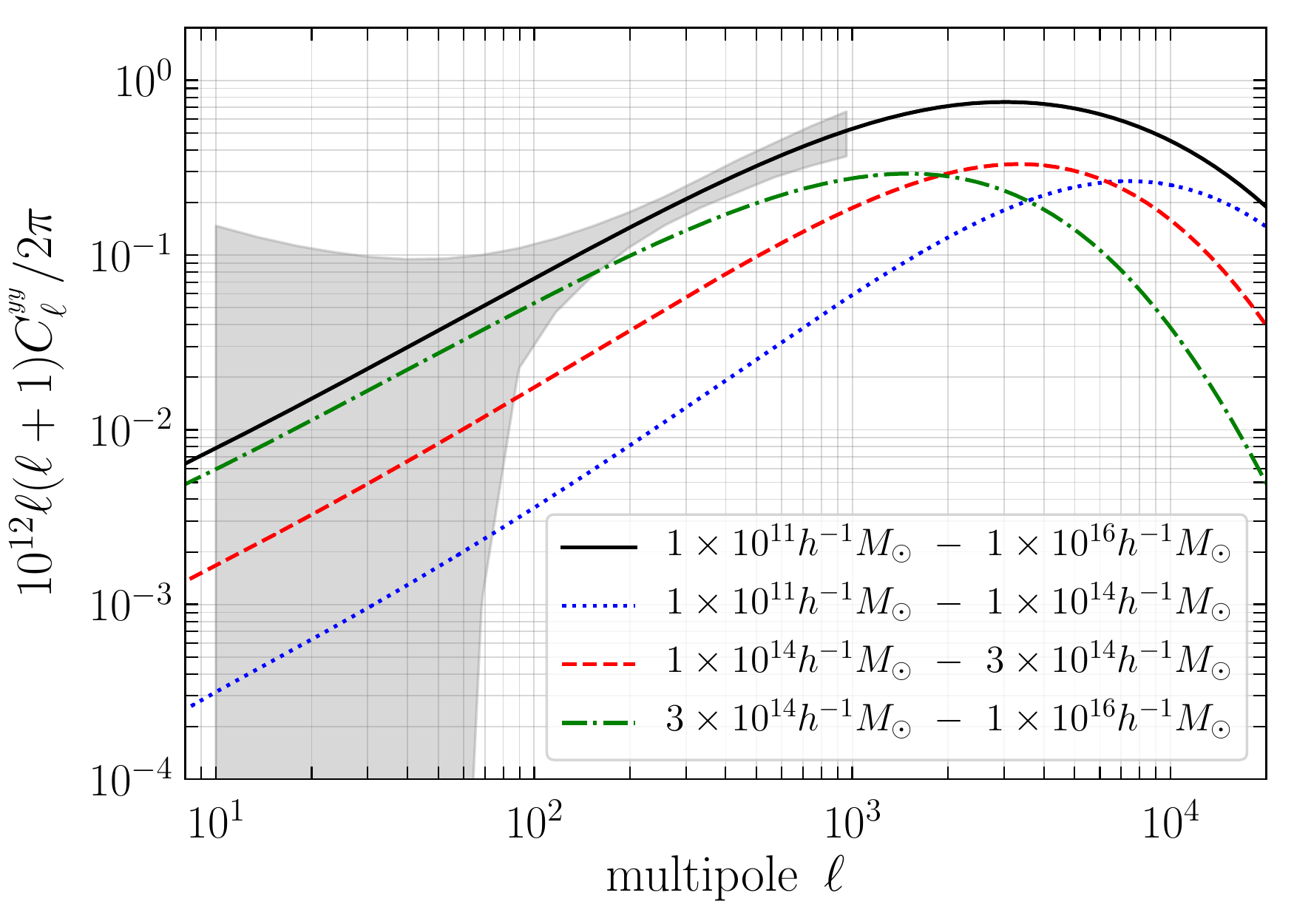}
\par\end{centering}
\caption{The $yy$-power spectrum computed with {\tt CLASS-SZ} \citep{Bolliet:2017lha, CLASSII}, for a spatially flat $\Lambda$CDM Universe with $\sigma_8 = 0.8$, $h=0.7$, $\Omega_{\rm b} = 0.05$, $\Omega_{\rm c} = 0.26$, $\tau=0.07$, $n_{\rm S}=0.96$, \citet{Tinker:2008ff} halo mass function interpolated at $M_{500c}$, and Planck 2013 pressure profile \citep{2013A&A...550A.131P}, with mass bias $B=1.41$ ($b\simeq 0.29$). 
The gray area indicates the 68\% CL interval after foreground marginalization, for the multipole range used in the {\it Planck} tSZ analysis \citep[see][]{Bolliet:2017lha}.
We also illustrate the separate contributions of low-, intermediate- and high-mass halos to the total $yy$-power spectrum. 
}
\label{fig:tSZ-Mass-cut}
\end{figure}

Although the first assumption is expected to have a smaller effect, both simplifications need to be revisited. When setting $Y(\nu)\simeq Y_0(\nu)$ one implicitly assumes that the temperature of the medium responsible for the $y$-parameter ($\rightarrow$ integrated pressure) fluctuations is non-relativistic ($\varv_{\rm th}/c\ll 10^{-3}$). However, the {\it Planck} power spectrum analysis is mostly sensitive to clusters with large masses $M\gtrsim \pot{3}{14} h^{-1}\,M_\odot$ (see Fig.~\ref{fig:tSZ-Mass-cut}), dominating at $\ell \simeq 10^2-10^3$, and hence to electrons with typical temperature\footnote{We used hydrostatic equilibrium expressions to estimate the cluster temperature \citep[e.g., see][]{Arnaud2005, 2007ApJ...668....1N, Erler2017}: $k \Te \simeq 5\,\keV\, \big[E(z)\,M_{500}/3\times 10^{14} h^{-1}\,M_\odot\big]^{2/3}$ with 
normalized Hubble factor $E(z)=\sqrt{\Omega_{\rm m} (1+z)^3 + \Omega_\Lambda}$.} $k\Te\gtrsim 5\,\keV$. A similar conclusion is reached by looking at fig.~11 of \citet{Refregier2000} and fig.~6 of \citet{Komatsu:2002wc}. This statement is further supported when considering SZ clusters detected by {\it Planck} at high significance. In this case, one obtains a sample-averaged cluster temperature of $k\Te^{\rm X}\simeq (6.91\pm 0.08)\,\keV$ \citep{Erler2017} using measured X-ray mass-temperature scaling relations \citep{Reichert2011}, and $k\Te^{\rm tSZ} \simeq 6^{+3.8}_{-2.9}\,\keV$ by stacking clusters \citep{Erler2017}. At temperatures $k\Te\gtrsim 3-5\,\keV$, relativistic corrections to the tSZ signal become relevant, and hence $Y(\nu)\neq Y_0(\nu)$. Consequently, this affects the {\it Planck} tSZ analysis, as we show here.

\begin{figure}
\begin{centering}
\includegraphics[width=0.98\columnwidth]{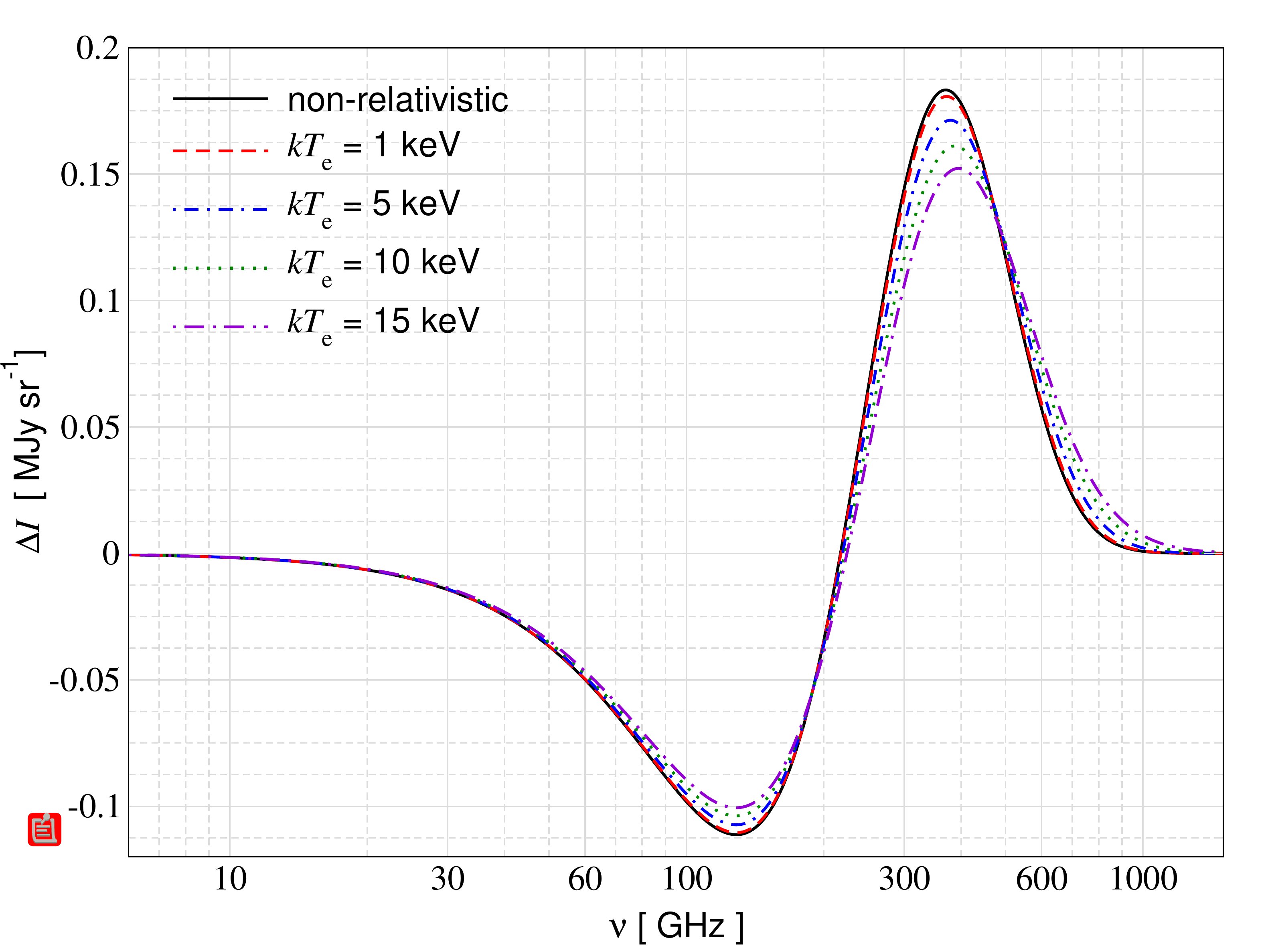}
\par\end{centering}
\caption{Illustration for the effect of relativistic temperature corrections on the tSZ signal, $\Delta I\equiv S(\nu, \Te)=y\,Y(\nu, \Te)$. At higher temperature, the tSZ signal broadens and shifts upward, leading to a reduction of the overall tSZ intensity at fixed Compton-$y$ parameter (we used $y=10^{-4}$).}
\label{fig:sed}
\end{figure}

Relativistic temperature corrections to the SZ signal can be accurately included using {\tt SZpack} \citep{Chluba2012SZpack}. Figure~\ref{fig:sed} illustrates the variations of the tSZ signal with the electron temperature. Relativistic corrections lead to a broadening of the tSZ intensity with systematic shift towards higher frequencies, reducing its overall amplitude at fixed Compton-$y$ parameter. This inevitably leads to an underestimation of $y$, if $Y_0(\nu)$ is used in the analysis. A similar conclusion was recently reached in \citet{Erler2017}, where the effect on the considered cluster sample was $\Delta y/y \simeq 7-14\%$. Hence, the amplitude of $C^{yy}_\ell$ is underestimated, an effect that propagates to the tSZ observables such as $\sigma_8$. This is further supported by the analysis of \citet{Hurier2017rSZ}. Similarly, relativistic tSZ should affect cluster number count statistics \citep{Planck2013SZ} and SZ analyses targeting neutrino masses and primordial non-Gaussianity \citep[e.g.,][]{Hill2013}.

\section{Formulation of the problem and results}
Using a tSZ temperature moment expansion \citep{Chluba2012moments} about pivot electron temperature, $\Teb$, we can express the tSZ signal, $S(\nu)=y\,Y(\nu, \Te)$, across the sky using the frequency-dependent spherical harmonic coefficients
\beal
S_{\ell m}(\nu) &\simeq Y(\nu, \Teb) \,y_{\ell m} +Y^{(1)}(\nu, \Teb) \,y^{(1)}_{\ell m}+ \frac{1}{2}Y^{(2)}(\nu, \Teb)\,y^{(2)}_{\ell m},
\end{align}
keeping terms up to second order in $\Delta \Te=\Te-\Teb$. For convenience, we introduced the derivatives $Y^{(k)}=\partial^k_T Y(\nu, T)$. We also defined the spherical harmonic coefficients, $y^{(k)}_{\ell m}=[(\Te-\Teb)^k y]_{\ell m}$, which generally each have different spatial morphology \citep{Chluba2012moments}. Assuming isothermal clusters, we furthermore have $y^{{\rm iso},(k)}_{\ell m}=(\Te-\Teb)^k y_{\ell m}$, an approximation that we will use below.

We still have to determine the pivot temperature $\Teb$ introduced above. One natural choice would be the average $y$-weighted SZ temperature, obtained by requiring $\big<y^{(1)}_{00}\big>\equiv \big<[(\Te-\Teb) y]_{00}\big>=0$, which yields $k\Teby\simeq\big<k[\Te y]_{00}\big>/\big<y_{00}\big>$. Within $\Lambda$CDM this has been estimated as $k\Teby\simeq 1.3\,\keV$ with all-sky $y$-parameter, $\left<y\right>\simeq \pot{2}{-6}$ \citep{Hill2015, abitbol_pixie}. This value for the average electron temperature is dominated by the contributions from low-mass halos ($M\lesssim \pot{{\rm few}}{13} h^{-1}\,M_\odot$).
However, for the tSZ power spectrum, a different weighting is relevant, which depends on details of the cluster mass function and temperature-mass relation. This increases the effective cluster sample temperature and hence the importance of relativistic corrections relevant to the tSZ power spectrum analysis, as we illustrate next.

To obtain the tSZ power spectrum, we have to compute the ensemble average $\big<S_{\ell m}^*\,S_{\ell m}\big>$. Because of isotropy and homogeneity, for a spherical cluster profile this yields $\big<y^*_{\ell m}\,y_{\ell m}\big>\rightarrow \big<|y_{\ell}|^2\big>$ \citep[e.g., see appendix of][for an explicit derivation]{Hill2013}, where $|y_{\ell}|^2$ is the 2D Fourier transform of the projected Compton $y$-parameter \citep[e.g.,][]{Komatsu:2002wc, Hill2013}. Again keeping only terms up to second order in $\Delta \Te$, with similar arguments we find the expansion of the theoretical tSZ power spectrum: 
\beal
\label{eq:C_ell}
C^{\rm tSZ}_\ell(\nu)&\simeq
Y^2(\nu, \Teb)\left<|y_\ell|^2\right>
+2Y(\nu, \Teb)\,Y^{(1)}(\nu, \Teb)\left<y^*_\ell\,y^{(1)}_\ell\right>
\\[1mm] \nonumber
&
+
\big[Y^{(1)}(\nu, \Teb)\big]^2
\left<|y^{(1)}_\ell|^2\right>
+
Y(\nu, \Teb)\,
Y^{(2)}(\nu, \Teb)
\left<y^*_\ell \,y^{(2)}_\ell\right>.
\end{align}
This expression shows that through relativistic corrections the tSZ power spectrum receives contributions from higher order statistics of the $y$-parameter and electron temperature fields.
These new terms are absent if $Y(\nu, \Te)\simeq Y_0(\nu)$ and lead to additional non-trivial frequency dependence. Similar effects were previously discussed for individual clusters \citep{Chluba2012moments}, but here we highlight the effects for ensembles of clusters. 

\begin{figure}
\begin{centering}
\includegraphics[width=\columnwidth]{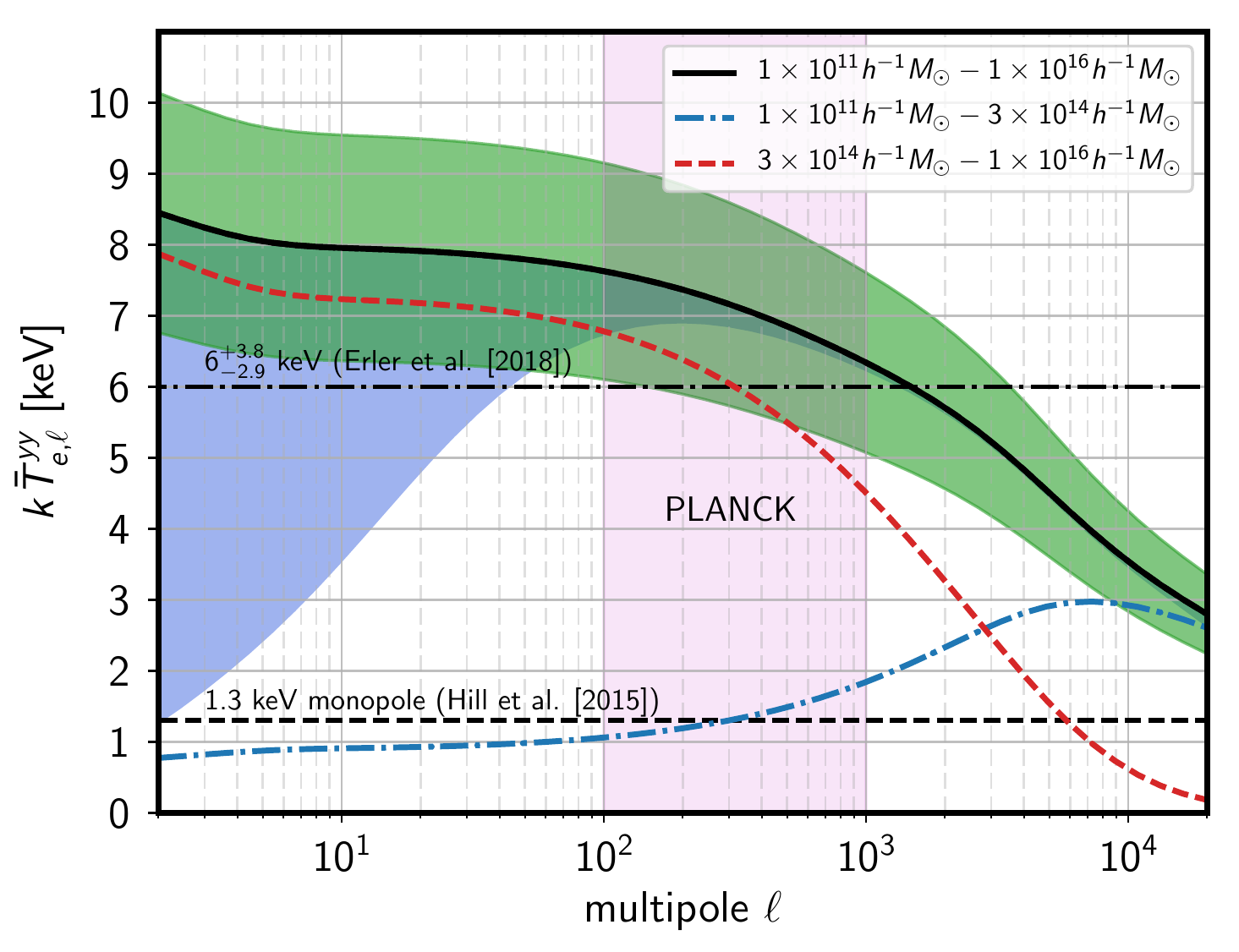}
\par\end{centering}
\caption{Effective $C^{yy}_\ell$-weighted electron temperature for different multipoles computed using {\tt CLASS-SZ} \citep{Bolliet:2017lha, CLASSII} with main settings like in Fig.~\ref{fig:tSZ-Mass-cut}. At high-$\ell$, low-temperature systems dominate, giving effective temperature $k\Tebell\simeq 2-3\,\keV$. Around $\ell\simeq 10^2-10^3$, which is most relevant to the {\it Planck} tSZ analysis, we find $k\Tebell\simeq 5-9\,\keV$. At $\ell\lesssim 10^2$, we obtain $k\Tebell\simeq 6-10\,\keV$. However, uncertainties in the assumed mass-temperature relation and its redshift dependence and the total amount of diffuse gas lead to large ambiguities (green and blue bands) that will have to be quantified more carefully.}
\label{fig:kTe_ell}
\end{figure}
In Eq.~\eqref{eq:C_ell}, we can now chose the pivot temperature, $\Teb$, to minimize contributions from higher order terms in $\Delta \Te$. In fact, this makes $\Teb$ a scale-dependent quantity, $\Tebell$, which can be defined by demanding $\big<y^*_\ell\,y^{(1)}_\ell\big>=0$ at each multipole $\ell$, cancelling the leading order correction term in Eq.~\eqref{eq:C_ell}. It is beyond the scope of this paper to include the spatial variations of the electron temperature within each cluster \citep[see][for some related discussion]{Chluba2012moments}. However, assuming an isothermal temperature profile for each cluster (i.e., $y^{(1)}_\ell=[\Te(M, z)-\Teb]\,y_\ell$), we find 
\beal
\label{eq:Te_ell}
k\Tebell = \frac{\big<k\Te(M, z)\,|y_\ell|^2\big>}{\big<|y_\ell|^2\big>} = \frac{C^{\Te, yy}_\ell}{C^{yy}_\ell}
\end{align}
to ensure $\big<y^*_\ell\,y^{(1)}_\ell\big>= 0$, such that in Eq.~\eqref{eq:C_ell} only second-order terms in $\Delta \Te$ remain. In the standard {\it Planck} analysis, $k\Tebell$ is arbitrarily set to zero. This choice biases the derived parameters since in this case higher order terms are not minimized. We remind the reader that the average $\big<\ldots\big>$ includes integrals over the cluster mass function and redshift. Following the formalism of \cite{Komatsu:2002wc}, the evaluation of $C^{\Te, yy}_\ell$ boils down to replacing $|\tilde{y}_\ell |^2$ by $k\Te(M, z)\,|\tilde{y}_\ell |^2$ in equation (1) of their work. 

One can think of Eq.~\eqref{eq:Te_ell} as a {\it $C^{yy}_\ell$-weighted temperature}. In Fig.~\ref{fig:kTe_ell}, we illustrate its scaling with multipole $\ell$ as obtained by modifying {\tt CLASS-SZ}. This highlights that at high-$\ell$/low-mass, colder systems dominate, yielding $k\Tebell \simeq 2-3\,\keV$. Around $\ell \simeq 10^2-10^3$, which is most relevant to the {\it Planck} tSZ analysis, we find an average temperature of $k\Tebell\simeq 5-9\,\keV$ for the $\Lambda$CDM cosmology. This estimate depends on the details assumed for the gas physics (e.g., the temperature-mass relation, feedback efficiencies and redshift scalings) that will have to be computed more carefully. These uncertainties are indicated by the green ($\pm 20\%$) band in Fig.~\ref{eq:Te_ell}. However, our halo-model calculations further justify our statements above, and suggest that $k\Teb\simeq 5\,\keV$ provides a {\it conservative} reference value. We also note that in the computations with {\tt CLASS-SZ} we only included contributions from the one-halo term, as the two-halo term is subdominant \citep[e.g.,][]{Hill2013}.

At large angular scales, the effective cluster temperature is expected to drop, approaching $k\Teby\simeq 1.3\,\keV$ found for the monopole\footnote{Note that $k\bar{T}^{yy}_{\rm e, \ell=0}$ is generally not expected to equal $k\Teby\simeq 1.3\,\keV$, which was computed using $k\Teby=\big<k\Te(M, z)\,y_0\big>/\big<y_0\big>$ ($y$-weighted temperature) as opposed to $k\bar{T}^{yy}_{\rm e, \ell=0}=\big<k\Te(M, z)\,y^2_0\big>/\big<y^2_0\big>$, which is relevant here.} \citep{Hill2015, abitbol_pixie}. This is due to the presence of diffuse, warm gas \citep[e.g.,][]{Hansen2005}, which should not contribute much to $C^{\Te, yy}_\ell$ but can increase $C^{yy}_\ell$ noticeably. Using \citet{Hansen2005}, we estimate this effect by adding $10^{12} \,\ell(\ell+1)\,C^{yy, \rm warm}_\ell /2\pi \lesssim 0.01$ for the warm diffuse component to $C^{yy}_\ell$. At large angular scales ($\ell\lesssim 10^2$), this contribution dominates and, in spite of large uncertainties, causes $k\Tebell$ to decline (see blue band in Fig.~\ref{fig:kTe_ell}).

A detailed study of all the associated effects on the tSZ power spectrum encoded by Eq.~\eqref{eq:C_ell} will be carried out in a follow-up paper. At leading order, the impact of relativistic SZ on the tSZ power spectrum can be captured by 
$C_\ell^{\mathrm{tSZ}}(\nu) \propto C_\ell^{yy} / f(\Teb)$, 
where $f(\Teb)$ generally is scale- and frequency-dependent. However, after component separation, which targets $C_\ell^{yy}$ not $C^{\rm tSZ}_\ell(\nu)$, we can assume one effective temperature at the current level of precision. For $C_\ell^{yy}\propto \sigma_8^{8.1}$ this implies that the {\it Planck} tSZ power spectrum analysis actually constrains $\sigma_8^\star \simeq \sigma_8/f(\Teb)^{1/8.1}$ when omitting relativistic corrections. Thus, the value for $\sigma_8$ obtained in the analysis is lowered by
\begin{equation}
\label{eq:bias_s8}
\Delta \sigma_8/\sigma^\star_8 \simeq f(\Teb)^{1/8.1}-1,
\end{equation}
with $\Delta \sigma_8\equiv \sigma_8-\sigma^\star_8$. We show that at the current level of precision this yields a systematic shift of  $\simeq 1\sigma$ towards larger $\sigma_8$ once relativistic corrections are included in the {\it Planck} tSZ power spectrum analysis. Since we do not know the exact value for $k\Teb$, this results in additional uncertainties that in the future will need to be quantified and marginalized over.

In the \emph{Planck} 2015 data analysis \citep{Planck2016ymap}, the Compton-$y$ map was estimated through a weighted linear combination of the frequency maps, with minimum variance to mitigate foreground contaminations. The weights assigned to each frequency map were determined to achieve unit response to the {\it non-relativistic} tSZ energy spectrum, $Y_0(\nu)$, thus ignoring relativistic corrections. In other words, it was implicitly assumed that the temperature of all clusters is $k\Te \ll 1\,\keV$ (cf. Fig.~\ref{fig:sed}), while here we argued that the average temperature of clusters relevant to the tSZ power spectrum analysis is $k\Te\gtrsim 5\,\keV$. 
We thus revised the estimation of the {\it Planck} tSZ $y$-map by modifying the \textsc{NILC} component separation algorithm \citep{Remazeilles2011,Remazeilles2013} that was adopted in \cite{Planck2016ymap}. We used the relativistic tSZ energy spectrum, $Y(\nu, \Te)$, for different temperatures $\Te>0$ instead of the non-relativistic spectrum to construct the \textsc{NILC} filters. Bandpass averaging had no large impact on the results, although at higher sensitivity this may not be the case. 

\begin{figure}
\begin{centering}
\includegraphics[width=\columnwidth]{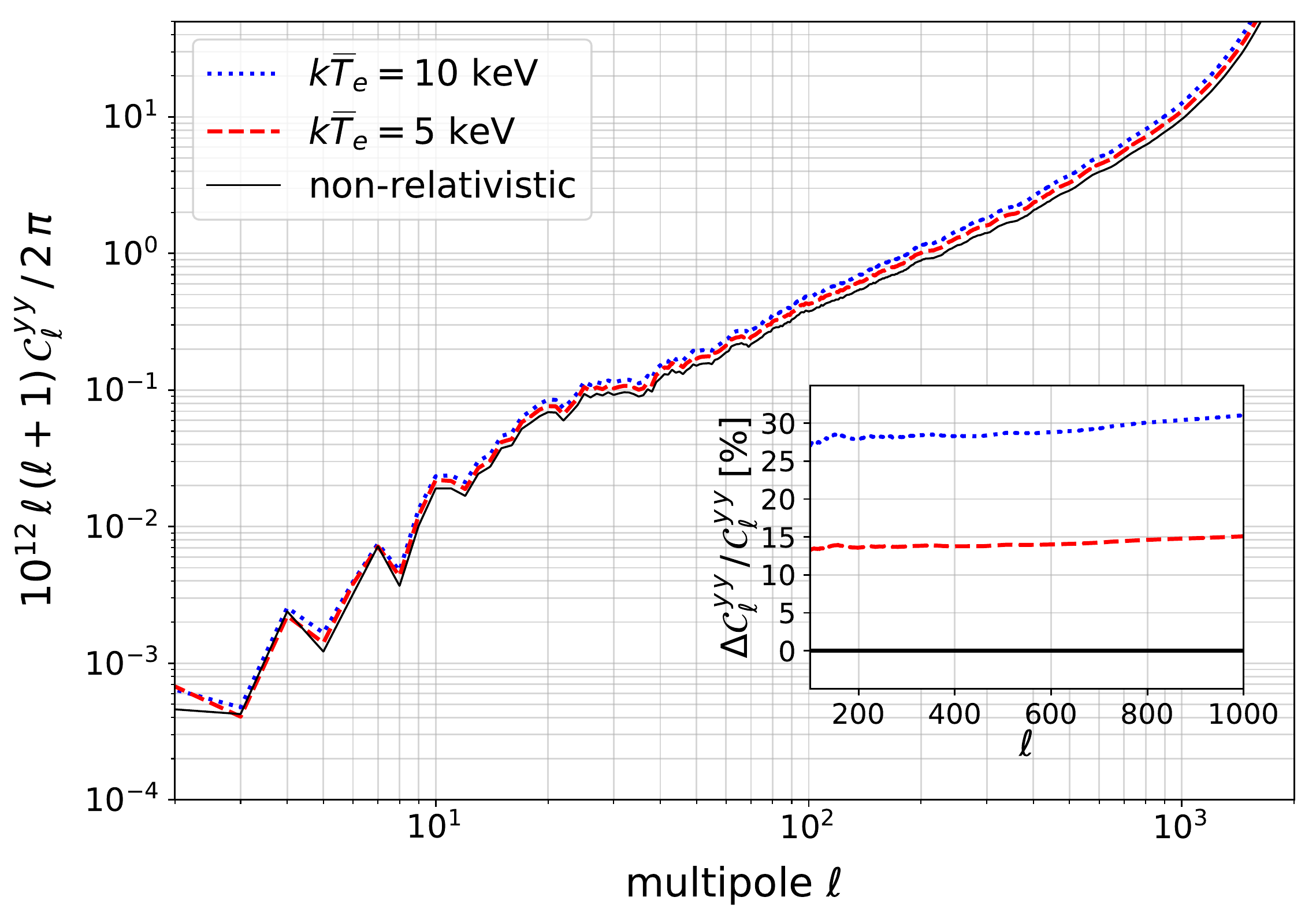}
\\[0mm]
\includegraphics[width=\columnwidth]{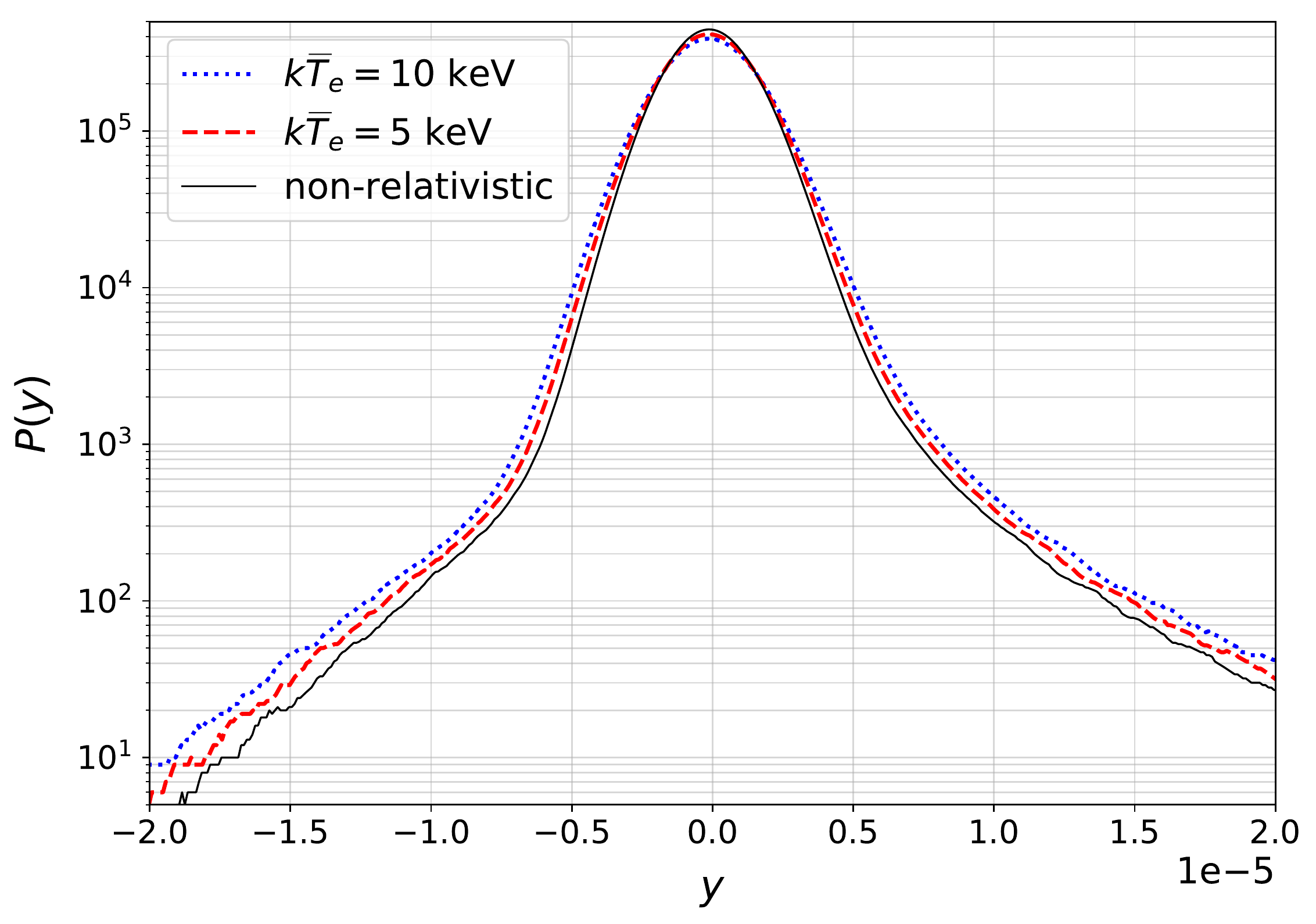}
\par\end{centering}
\caption{\emph{Upper panel}: angular power spectrum of the \textsc{NILC} $y$-maps on $80$\% of the sky from \emph{Planck} data obtained using different effective electron temperatures, $k\Teb$. The inlay highlights the relative difference in comparison with the non-relativistic case. {\it Lower panel}: corresponding one-point PDF of the $y$-maps on $40$\% of the sky.}
\label{fig:ps}
\end{figure}

We applied our revised \textsc{NILC} filters to the {\it Planck} 2015 data, assuming $k\Teb=5$ and $10\,\keV$, to reconstruct the tSZ $y$-map. We then estimated $C^{yy}_\ell$ and the one-point PDF after foreground marginalization from the obtained $y$-map, as presented in Fig.~\ref{fig:ps}. The amplitude of the tSZ power spectrum increases noticeably with $\Te$, as anticipated. Similarly, the width and skewness of the PDF are modified. By comparing our results to those obtained using the non-relativistic tSZ energy spectrum we find
\bsub
\beal
\label{eq:Power_analysis}
f(\Te)&\simeq C^{yy}_\ell(\Te) / C^{yy}_\ell(\Te=0) \simeq 1 + 0.15 \,\Bigg[\frac{k\Te}{5\keV}\Bigg]
\\
g(\Te)&\simeq \mathcal{S}(\Te) / \mathcal{S}(\Te=0) \simeq 1 + 0.28 \,\Bigg[\frac{k\Te}{5\keV}\Bigg]
\end{align}
\esub
to represent the changes of the power spectrum amplitude and skewness of the one-point PDF, $\mathcal{S}=\langle y^3\rangle$. The result for $f(\Te)$ can also be estimated by comparing the amplitude of $Y_0(\nu)$ and $Y(\nu, k\Te=5\,\keV)$ in the $\nu=353\,\GHz$ channel of {\it Planck}, yielding $f(5\,\keV) \simeq Y^2_0(\nu)/Y^2(\nu, 5\,\keV)\simeq 1.19$. In \citet{Erler2017}, it was found that after foreground marginalization the $\nu=353$ and $143\,\GHz$ channels were indeed driving the constraints on relativistic tSZ. This is related to the ability of {\it Planck} to distinguish foregrounds from the signal, indicating that more careful simulations are needed to quantify the effect.

Various estimates of $\sigma_8$ exist in the literature \citep[see][for references]{Planck2016ymap, Bolliet:2017lha}. Typical central values are $\sigma_8\simeq 0.8$ with $1\sigma$ error $\simeq 0.02$. With Eq.~\eqref{eq:bias_s8} and Eq.~\eqref{eq:Power_analysis}, we can directly write
\begin{equation}
\label{eq:bias_s8_result}
\Delta \sigma_8/\sigma^\star_8\simeq 0.019 \,\left[\frac{k\Teb}{5\,\keV}\right]
\end{equation}
for the systematic shift expected in the tSZ power spectrum analysis due to relativistic corrections. Assuming a fiducial value $\sigma_8\simeq 0.8$ yields $\Delta \sigma_8 \simeq 0.015 \,\big[k\Teb/5\,\keV \big]$, which is comparable to the current $1\sigma$ uncertainty on $\sigma_8$. From the skewness we find $\Delta \sigma^{\mathcal{S}}_8/\sigma^\star_8\simeq 0.025 \,\big[k\Teb/5\,\keV\big]$, implying $\Delta \sigma^{\mathcal{S}}_8 \simeq 0.02\, \big[k\Teb/5\,\keV\big]$, in good agreement with Eq.~\eqref{eq:bias_s8_result}.

In the {\it Planck} 2015 analysis of the Compton-$y$ map, the collaboration reported constraints on $\sigma_8$ that are in mild tension with the CMB anisotropy constraints, with the tSZ analysis yielding systematically lower values \citep{Planck2016ymap}. A detailed review of the various results and their differences is beyond the scope of this paper, but from Eq.~\eqref{eq:bias_s8_result} it follows that {\it all} the tSZ power-spectrum-derived constraints on $\sigma_8$ are currently biased low by about $1\sigma$. This means that including relativistic temperature corrections could alleviate the tension with the CMB anisotropy data. To reduce the tension to below $1\sigma$, $\Delta \sigma_8\simeq 0.03-0.05$ is required, implying $k\Teb\simeq 10-15\,\keV$. This seems quite high, since only the most massive clusters seen in our Universe reach comparable temperatures \citep{ElGordo2012, Chluba2012moments}. However, relativistic corrections play a part in the story, already adding to the total error budget at the current level of precision.

As outlined by a number of recent works \citep[e.g.,][]{Hurier:2017jgi,Salvati:2017rsn,Makiya:2018pda,Bolliet:2018yaf}, the tension between tSZ probes and CMB temperature anisotropy can be rephrased in terms of the mass bias, $B=1/(1-b)$, rather than $\sigma_8$. Hydrodynamical simulations suggest $b\simeq 0.2$ or $B\simeq 1.25$ \citep[e.g.,][]{2016MNRAS.455.2936S}. This can arise due to departure from hydrostatic equilibrium (e.g., non-thermal pressure); however, other effects such as systematics in the X-ray mass calibration also contribute \citep[e.g., see][for discussions]{2007ApJ...655...98N, 2009ApJ...705.1129L, Shaw2010SZ, Shi:2014msa, Henson2017}. 

A more practical approach needs to take the uncertainty in the mass bias into account. Given the current tSZ constraint on $F=\sigma_8 \big(\Omega_{{\mathrm{m}}}/B\big)^{0.40}\,h^{-0.21}$ \citep{Bolliet:2017lha}, one finds  $B=1.58\pm 0.13$ (68\% CL) with CMB $TT$ + lensing, i.e., $b=0.37\pm0.05$ (68\% CL). Accounting for relativistic tSZ, the mass bias is driven towards lower values, more consistent with hydrodynamical simulations. Indeed the constraint on $F$ should be revised to $F^\star =F/f(\Te)^{1/8.1}$, implying $\Delta b\simeq 0.046\,(1-b)\big[k\Te/5\,\keV\big]$. We also highlight that $k\Tebell$ defined by Eq.~\eqref{eq:Te_ell} is relatively insensitive to mass-bias parameter, $b$, but depends on $\sigma_8$ as ${k\Tebell \simeq 7\,\keV\,(\sigma_8/0.8)^{2}}$ (at $\ell\simeq 10^2-10^3$), thus in principle offering a new way to break parameter degeneracies. We will explore this idea in the future.

\section{Conclusions}
To summarize, we took an important first step towards including the effects of relativistic temperature corrections on tSZ power spectrum analyses, providing a new formalism for capturing the associated effects, i.e., Eq.~\eqref{eq:C_ell}. Applying the method to {\it Planck}, we showed that this can help reduce part of the tension between different cosmological probes of $\sigma_8$. However, it will be important to directly estimate the average electron temperature, $k\Teb$, which has large uncertainties that need to be marginalized over. For example, cluster gas physics and feedback processes affect the temperature-mass relation and its redshift evolution. It is also clear that existing temperature estimates (e.g., X-ray/spectroscopic versus mass and $y$-weighted temperatures) differ significantly \citep[e.g.,][]{Kay2008}, demanding further quantification. In addition, at large angular scales, contributions from the diffuse, relatively cold gas cannot be ignored \citep[e.g.,][]{Zhang2004, Hansen2005, Hill2015}. Degeneracies with CMB foregrounds at different scales will also have to be studied more carefully. 

We highlighted that the {\it shape} of tSZ power spectrum depends on higher order statistics of the $\Te$ and $y$ fields (see Eq.~\ref{eq:C_ell}) in a {\it frequency-dependent manner}. This is caused by weighted averages of spatially varying spectral energy distributions. Similar ideas have recently been discussed in connection with CMB foreground analyses \citep{Chluba2017foregrounds}. This 
opens a new window for exploring the statistical and physical properties of clusters in our Universe. Extracting these signals will require high sensitivity and broad spectral coverage, as discussed for space mission concepts like {\it CORE} \citep{Delabrouille2018, Remazeilles2018},  {\it LiteBIRD} \citep{Suzuki2018}, {\it PICO} and {\it CMB-Bharat}. 

Finally, in this paper, as an example we highlighted the effects on the tSZ power spectrum and connections to $\sigma_8$. Relativistic corrections will also be relevant to tSZ constraints on the sum of neutrino masses and potentially primordial non-Gaussianity. They are furthermore expected to affect cluster number counts in a similar manner, increasing the number of clusters at a given signal-to-noise ratio threshold \citep[see][for some related discussion]{Fan2003}. The refinements discussed here will also become important for the next-generation CMB experiments such as Simons Observatory, CMB-S4 \citep{Abazajian2016S4SB} and CCAT-prime \citep{CCATp2018}, providing new science targets related to cluster astrophysics and their impact on cosmological observables.

\smallskip
\small
{\it Acknowledgments.} We cordially thank Kaustuv Basu, Richard Battye, Jens Erler, Colin Hill, Eiichiro Komatsu and Fabian Schmidt for valuable comments on the manuscript. 
We also thank the anonymous referee for suggestions.
This work was supported by the ERC Consolidator Grant CMBSPEC (No. 725456). JC is supported by the Royal Society as a Royal Society University Research Fellow at the University of Manchester, UK. This analysis is partly based on observations obtained with Planck (http://www.esa.int/Planck), an ESA science mission with instruments and contributions directly funded by ESA Member States, NASA, and Canada. 

\small 
\vspace{-3mm}
\bibliographystyle{mn2e}
\bibliography{main,Lit}

\end{document}